# Scalable intensity-based photonic matrix-vector multiplication processor using single-wavelength time-division-multiplexed signals

CHENGLI CHAI,[1] RUI TANG,[1,*] MAKOTO OKANO,[2] KASIDIT TOPRASERTPONG,[1] SHINICHI TAKAGI,[1] AND MITSURU TAKENAKA[1]

[1]*Department of Electrical Engineering and Information Systems, The University of Tokyo, Tokyo 113-8656, Japan*
[2]*National Institute of Advanced Industrial Science and Technology, Ibaraki 305-8568, Japan*
*\*ruitang@mosfet.t.u-tokyo.ac.jp*

**Photonic integrated circuits provide a compact platform for ultrafast and energy-efficient matrix-vector multiplications (MVMs) in the optical domain. Recently, schemes based on time-division multiplexing (TDM) have been proposed as scalable approaches for realizing large-scale photonic MVM processors. However, existing demonstrations rely on coherent detection or multiple wavelengths, both of which complicate their operations. In this work, we demonstrate a scalable TDM-based photonic MVM processor that uses only single-wavelength intensity-modulated optical signals, thereby avoiding coherent detection and enabling simplified operations. A 32-channel processor is fabricated on a Si-on-insulator (SOI) platform and used to experimentally perform convolution operations in a convolutional neural network (CNN) for handwritten digit recognition, achieving a classification accuracy of 93.47% for 1500 images.**

**Introduction.** The increasing size of deep learning models has made power consumption a major concern, with matrix multiplications being identified as a significant computational bottleneck. Although graphics processing units (GPUs) are widely used to enhance computational speed, the pace of improvement in energy efficiency has slowed as Moore's law nears its physical limit. This has driven the development of energy-efficient hardware accelerators for deep learning. Photonic neural networks (PNNs) have emerged as a promising solution, enabling analog matrix-vector multiplications (MVMs) on compact photonic integrated circuits with exceptional speed and energy efficiency [1–5]. Various architectures for photonic MVM have been proposed, including coherent architectures that use both the amplitude and phase information of light [6–9], and intensity-based architectures that use only the amplitude information [10–16]. However, the scalability of these architectures is rather limited due to the need for a large number of optical modulators—typically $N^2$ optical modulators are required to represent an $N \times N$ matrix. Recently, scalable architectures based on time-division multiplexing (TDM) have been proposed [17–23], in which the number of optical modulators is significantly reduced from $N^2$ to $N$ for the same matrix size. Despite this progress, the operations of existing TDM-based devices are still relatively complicated due to the need for coherent detection or multiple wavelengths [18,19,23], and large-scale demonstrations are still lacking.

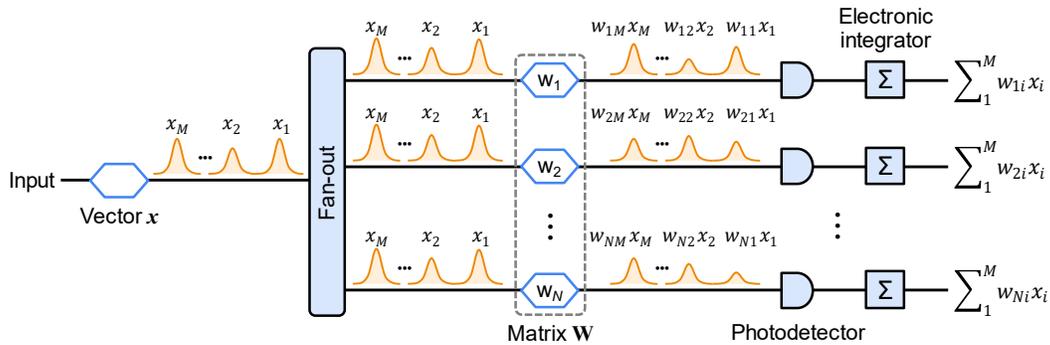

Fig. 1. Operation principle of this scheme. A single-wavelength input light is modulated by an optical intensity modulator, which generates an $M$-element vector $x$ by sequentially encoding each element onto the light intensity. These TDM signals are equally split into $N$ channels and further modulated by $N$ modulators, each generating one row in an $N \times M$ matrix $\mathbf{W}$. The two intensity modulations perform the multiplication between the vector and matrix elements. These signals are detected by photodetectors, generating photocurrents proportional to the optical power, which are then integrated by electronic integrators to perform the accumulation operation.

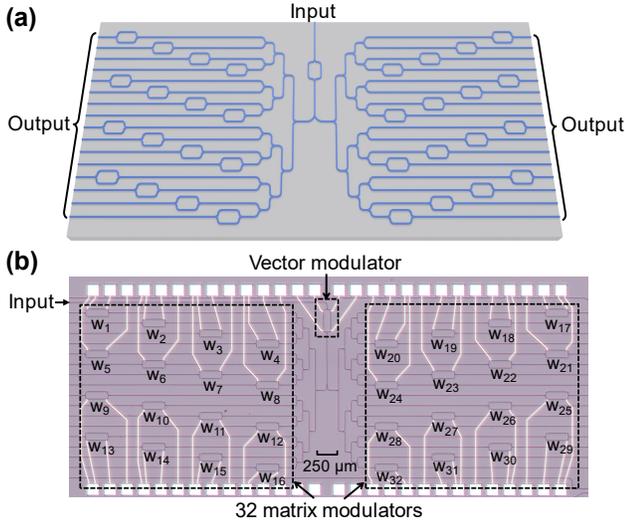

Fig. 2. (a) Schematic structure of the 32-channel processor. (b) A microscope image of the 32-channel processor fabricated on an SOI platform.

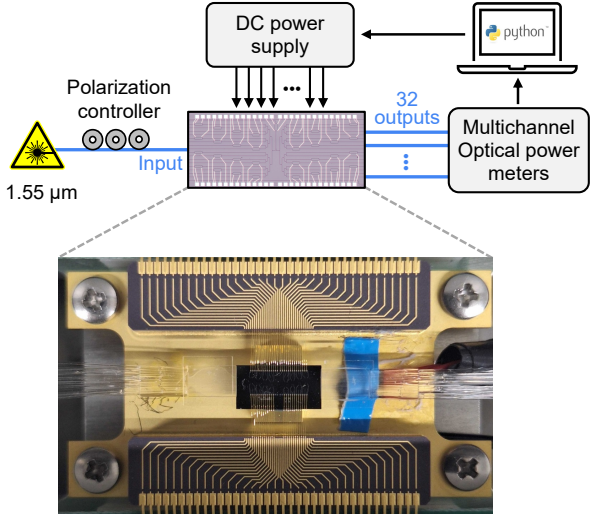

Fig. 3. Experimental setup and an image of the packaged chip. The chip is wire-bonded for electrical connections and packaged with two fiber arrays.

In this work, we propose and demonstrate a scalable TDM-based photonic MVM processor that uses only single-wavelength intensity-modulated optical signals, which eliminates the need for coherent detection and simplifies operations. A large-scale ($N=32$) processor is fabricated on a Si-on-insulator (SOI) platform and used to perform convolution operations in a convolutional neural network (CNN) for handwritten digit recognition, achieving a classification accuracy of 93.47%.

**Results.** The operation principle of this scheme is illustrated in Fig. 1. A single-wavelength optical input is modulated using an optical intensity modulator, which generates an $M$-dimensional vector $x$ by sequentially encoding each element onto the intensity of the optical signal. Note that $M$ can be arbitrary. These TDM signals are evenly distributed among $N$ channels and subsequently modulated using $N$ modulators, each corresponding to a row of the $N \times M$ weight matrix $\mathbf{W}$. The two stages of intensity modulation perform the element-wise multiplication of the vector and matrix. The twice-modulated optical signals are converted into electrical signals by photodetectors, generating photocurrents that are proportional to the optical powers, which are subsequently integrated by electronic integrators for the accumulation operation [17]. Here, high-speed intensity modulators are desired, and the vector and matrix modulators should be synchronized to function properly.

To demonstrate this concept, a 32-channel ($N=32$) processor is fabricated on an SOI platform by a commercial foundry (Applied Nanotools) using electron-beam lithography. The schematic structure and a microscope image of the processor are shown in Fig. 2. The Si waveguides have a standard core size of $500 \times 220$ nm$^2$ and a typical propagation loss of 1.2 dB/cm. Edge couplers based on inverse tapers are used to couple light into and out of the chip. Cascaded stages of 1×2 multimode interference (MMI) couplers are used to split the light into 32 channels. Tunable Mach-Zehnder interferometers (MZIs) are employed as intensity modulators, using thermo-optic phase shifters (length: 220 μm, width: 4 μm) with a power consumption of less than 30 mW/π. Since the operation speed of thermo-optic phase shifters is not sufficiently fast, they are used here only for proof-of-concept purposes. High-speed electro-optic modulators or electro-absorption modulators should be employed in practical scenarios [9,24,25]. While photodetectors are not integrated on this chip due to platform limitations, they can be easily integrated onto the same chip using other foundry services [26]. The electronic integrator can be implemented as described in [18]. The chip is wire-bonded for electrical connections and packaged with two fiber arrays, with coupling losses ranging from 2.7 to 4.5 dB per coupling. Figure 3 shows the experimental setup and the packaged chip. Continuous-wave light at a wavelength of 1.55 μm is injected into the chip after its polarization is adjusted to the transverse electric (TE) mode. The chip temperature is stabilized at room temperature using a thermoelectric cooler. All phase shifters on the chip are driven by a 40-channel direct current (DC) power supply (NicsLab, XDAC-40MUB-R4G8). Optical signals from the 32 output ports are detected by two multi-channel optical power meters (Santec, OP760).

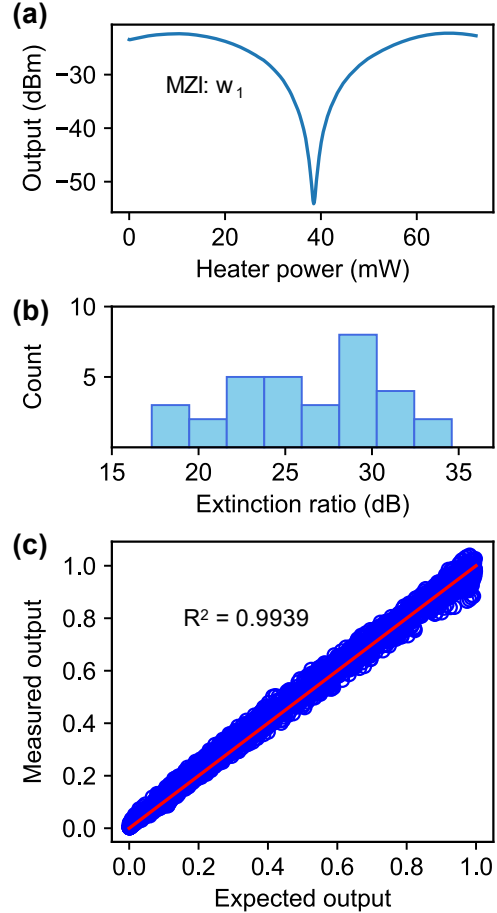

Fig. 4. (a) Characterization result of one MZI ($w_1$) when sweeping the electric power applied to the thermo-optic phase shifter, exhibiting an extinction ratio of 31.9 dB. (b) Measured extinction ratios for all 32 matrix MZIs. (c) Normalized measured output powers and expected output powers when 100 sets of random configurations are applied to all MZIs. A total of 3200 points are plotted, showing a high determination coefficient ($R^2$) of 0.9939.

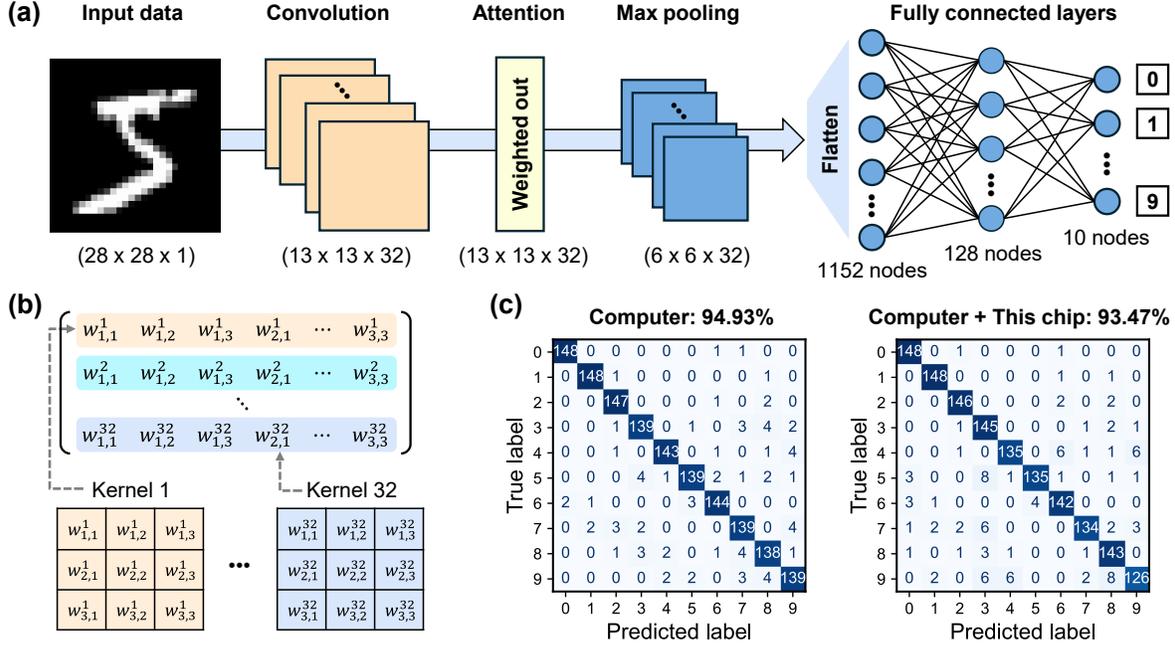

Fig. 5. (a) A CNN constructed for handwritten digit recognition, which consists of a convolution layer, an attention layer, a max pooling layer, and fully connected layers. (b) Converting 32 kernels into a 32×9 matrix for implementation on the photonic processor. Each 3×3 kernel is flattened into a row in the matrix. (c) Classification results for 1500 MNIST images when the inference is performed using a computer alone (left) and when the 32-channel processor is used to perform the convolution operations (right).

Each MZI is characterized to establish a lookup table that maps the electric power applied to the phase shifter to the normalized MZI transmittance. The result for one MZI is shown in Fig. 4(a), exhibiting an extinction ratio of 31.9 dB. The measured extinction ratios of all matrix MZIs are plotted in Fig. 4(b), with a mean value of 26.1 dB. Random initial phases are observed among the MZIs due to fabrication imperfections. Once lookup tables were established for all MZIs, we applied 100 sets of random configurations across all MZIs and measured the optical power at each output port. The measured powers were normalized and compared with the expected values, as shown in Fig. 4(c), which contains 3200 points in total. The determination coefficient ($R^2$) is as high as 0.9939, indicating the high operation fidelity of the fabricated chip.

A CNN for handwritten digit recognition is constructed, as shown in Fig. 5(a). This CNN consists of a convolution layer, an attention layer, a max pooling layer, and fully connected layers. It is trained on the MNIST dataset, which contains 60000 training images and 10000 test images. Each image is a single-channel grayscale image with 28×28 pixels. The convolution layer uses 32 kernels (kernel size: 3×3, stride: 2, padding: 0), which convert an input image into 13×13×32 feature maps. All kernel elements are constrained between 0 and 1 during training to enable implementation on the photonic processor. These feature maps then pass through a simple attention layer [27], which computes attention scores and generates a weighted output. The attention scores are stored in a 13×13 matrix **S**, calculated as

$$\mathbf{S} = \text{Softmax}(\mathbf{I}_{13\times13\times32}\mathbf{A}_{32\times1}), \quad (1)$$

where $\mathbf{I}_{13\times13\times32}$ represents the input feature maps, $\mathbf{A}_{32\times1}$ represents a trainable weight vector, and Softmax represents the softmax function. $\mathbf{A}_{32\times1}$ is optimized automatically during training. The weighted output is then generated as

$$\mathbf{O} = \mathbf{I}_{13\times13\times32} \odot \mathbf{S}_{13\times13}, \quad (2)$$

where $\odot$ represents the Hadamard product. The output is next processed through a max pooling layer (pool size: 2×2) and flatten into a vector, which is then processed by fully connected layers. Dropout layers are applied after the pooling layer and the first fully connected layer during training, with dropout probabilities of 0.25 and 0.5, respectively. The dropout layers are deactivated during inference. The rectified linear unit (ReLU) and softmax function are used as the nonlinear activators for the first and second fully connected layers, respectively. This CNN is trained on a computer using the Adam optimizer and the categorical cross-entropy loss function.

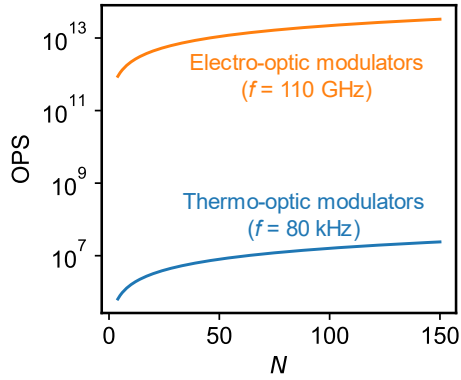

Fig. 6. Estimated computation speeds for this scheme under varying modulator types and matrix scales.

The 32-channel processor is used to perform the convolution operations by converting the 32 kernels into a 32×9 matrix, as illustrated in Fig. 5(b). Each 3×3 kernel is flattened into a row in the matrix, and each convolution region in the input image is flattened into a 9×1 vector. Thus, the convolution operations are executed as the multiplication of a 32×9 matrix and 9×1 vectors. For simplicity, the accumulation operation is performed on a computer by directly summing the measured outputs. The classification results for 1500 images are shown in Fig. 5(c). When using the computer alone, the classification accuracy is 94.93%. By performing the convolution operations on the photonic chip, a classification accuracy of 93.47% is experimentally achieved, further demonstrating the high operation fidelity of this chip.

The computation speed in operations per second (OPS) for this scheme is given by
$$\text{OPS} = 2fN, \quad (3)$$
where $f$ is the clock frequency, and the factor of 2 accounts for simultaneous multiplication and accumulation operations within one modulation cycle. For this chip, assuming $f$ is 80 kHz, which corresponds to a switching time of 12.5 μs [28], the computation speed is $5.12 \times 10^6$ OPS. The estimated computation speeds at various clock frequencies and matrix scales are shown in Fig. 6. Compact intensity modulators with 3-dB bandwidths exceeding 110 GHz have been demonstrated on both SOI and thin-film lithium niobate (TFLN) platforms [29,30]. If such modulators are used, a high computation speed of $2.82 \times 10^{13}$ OPS can be achieved with $N$=128.

**Conclusion.** We have demonstrated a scalable, intensity-based photonic MVM processor using single-wavelength TDM signals. A 32-channel processor was fabricated on an SOI platform and used to perform convolution operations in a CNN for handwritten digit recognition, achieving a classification accuracy of 93.47% for 1500 images. By further replacing thermo-optic modulators with high-speed electro-optic modulators, high-speed and energy-efficient MVMs can be performed using this architecture.

**Funding.** Japan Science and Technology Agency (CREST, JPMJCR2004); Japan Society for the Promotion of Science (22K14298).

**Disclosures.** The authors are aware of a recent preprint that demonstrates a similar architecture on a thin-film lithium niobate platform [25].

**Data availability.** Data underlying the results presented in this paper are available from the corresponding author upon reasonable request.